\documentclass[sn-aps]{sn-jnl} 
\catcode`\|=12\relax
\usepackage{orcidlink}
\usepackage[T1]{fontenc}
\usepackage{graphicx}
\usepackage{tikz} 
\usepackage[]{framed}
\usepackage{amsmath}
\usepackage{adjustbox}
\usepackage{xspace}
\usepackage{soul}
\usepackage{placeins}  
\usepackage{tcolorbox}
\tcbuselibrary{breakable}
\usepackage{longtable}
\usepackage{float}
\usepackage[inline]{enumitem}
\usepackage{booktabs}
\usepackage{array, multirow, multicol}
\usepackage[normalem]{ulem}
\usepackage{geometry}
\usepackage{tablefootnote}
\usepackage{threeparttable}

\newcommand\mmarco{\textsc{mMARCO}\xspace}
\newcommand\msmarco{\textsc{MS MARCO}\xspace}
\newcommand\clerc{\textsc{CLERC}\xspace}
\newcommand\ulyssesrfcorpus{\textsc{Ulysses-RFCorpus}\xspace}
\newcommand\quati{\textsc{Quati}\xspace}
\newcommand\juristcu{\textsc{JurisTCU}\xspace}
\newcommand\tcuvotes{\textsc{TCU Votes}\xspace}
\newcommand\stjjudgments{\textsc{STJ Judgments}\xspace}

\tcbset{
    style_prompt/.style={
        fontupper=\small
    }
}

\begin{document}

\title[\juristcu: A Brazilian Portuguese IR Dataset with QRELS]{\juristcu: A Brazilian Portuguese Information Retrieval Dataset with Query Relevance Judgments}

\author*[1]{\fnm{Leandro} \sur{Carísio Fernandes}\orcidlink{0000-0002-4114-2334}}\email{carisio@gmail.com}
\author[2]{\fnm{Leandro} \sur{dos Santos Ribeiro}\orcidlink{0009-0008-3715-9927} \email{leandro.santos.r@gmail.com}}
\author[2]{\fnm{Marcos} \sur{Vinícius Borela de Castro} \orcidlink{0009-0000-8633-1385} \email{borela@tcu.gov.br}}
\author[2]{\fnm{Leonardo} \sur{Augusto da Silva Pacheco}\orcidlink{0000-0001-6506-2366} \email{leonardo3108@gmail.com}}
\author[2]{\fnm{Edans} \sur{Flávius de Oliveira Sandes}\orcidlink{0009-0004-4553-4783} \email{edansfs@tcu.gov.br}}

\affil[1]{\orgname{Câmara dos Deputados}, \orgaddress{\city{Brasília}, \country{Brazil}}}
\affil[2]{\orgname{Tribunal de Contas da União (TCU)}, \orgaddress{\city{Brasília}, \country{Brazil}}}

\abstract{This paper introduces \juristcu, a Brazilian Portuguese dataset for legal information retrieval (LIR). The dataset is freely available\footnote{\url{https://huggingface.co/datasets/LeandroRibeiro/JurisTCU}}\hspace{0.3em} and consists of 16\,045 jurisprudential documents from the Brazilian Federal Court of Accounts, along with 150 queries annotated with relevance judgments. It addresses the scarcity of Portuguese-language LIR datasets with query relevance annotations. The queries are organized into three groups: real user keyword-based queries, synthetic keyword-based queries, and synthetic question-based queries. Relevance judgments were produced through a hybrid approach combining LLM-based scoring with expert domain validation. We used \juristcu in 14 experiments using lexical search (document expansion methods) and semantic search (BERT-based and OpenAI embeddings). We show that the document expansion methods significantly improve the performance of standard BM25 search on this dataset, with improvements exceeding 45\% in P@10, R@10, and nDCG@10 metrics when evaluating short keyword-based queries. Among the embedding models, the OpenAI models produced the best results, with improvements of approximately 70\% in P@10, R@10, and nDCG@10 metrics for short keyword-based queries, suggesting that these dense embeddings capture semantic relationships in this domain, surpassing the reliance on lexical terms. Besides offering a dataset for the Portuguese-language IR research community, suitable for evaluating search systems, the results also contribute to enhancing a search system highly relevant to Brazilian citizens.}


\maketitle

\section{Introduction}

Information retrieval (IR) deals with the representation, storage, organization, and retrieval of information to help users find relevant content within a collection of documents \cite{modern_ir}. Typically, a user formulates a query to express their information need and searches within an unstructured collection of documents, usually in text format \cite{manning2009introduction}.

One subfield of IR is Legal Information Retrieval (LIR). In this case, the document corpus includes judicial decisions, legislation, case law, contracts, jurisprudence, and other legal texts, which are often written in specialized language \cite{concept_lir}. Its importance lies not only in supporting legal professionals, researchers, and policymakers but also in enabling citizens to exercise their rights and engage with the legal system, which is essential in democratic societies.

The development of LIR systems presents some challenges. The volume of published legal documents is constantly increasing, and jurisprudence evolves over time, changing the prevailing interpretation of some legal application. Therefore, there is a continuous need to improve legal search systems, either by refining existing methods or exploring new search strategies.

Research on advancing IR systems relies on a standardized evaluation approach, using datasets that include a corpus (a collection of searchable documents), a fixed set of queries, and relevance judgments (qrels) indicating which documents are relevant to each query \cite{elastic_search_relevance}. Although there are some datasets available in Brazilian Portuguese, there is a scarcity of LIR datasets. Furthermore, to the best of our knowledge, there are no LIR datasets in Portuguese that include typical queries used in search systems with query relevance judgments. To contribute to research in this area, we present \juristcu, a dataset containing 16\,045 documents and 150 queries, each linked to 15 relevance-annotated documents, totaling 2\,250 judgments.

The \juristcu corpus consists of jurisprudence from the Brazilian Federal Court of Accounts (Tribunal de Contas da União – TCU). The TCU is responsible for auditing the efficient use of Brazilian public funds and provides access to over 500\,000 documents generated since 1992 that contain precedents on how legislation has been applied in various cases. These documents, grounded in legal norms and interpretations developed over time, form a body of jurisprudence that reflects the institution's normative understanding.

To facilitate access to these precedents, the TCU curates a smaller collection of documents, creating a "selected jurisprudence" dataset that contains deliberations chosen by a specialized team based on criteria of jurisprudential relevance. These curated documents include the Court's most important decisions and are the basis of \juristcu.

Besides presenting the dataset, we also conduct 14 experiments using lexical and semantic search techniques to establish benchmarks for it.

This paper contributes to the information retrieval research community focused on the Portuguese language (through the dataset and benchmark experiments) and also benefits the TCU itself, by providing an evaluation dataset to improve its search system for the benefit of Brazilian citizens.

The paper is organized into five sections: Section 2 reviews related IR and LIR datasets. Section~3 explains the methodology for creating the \juristcu dataset. Next, we use the dataset in a series of information retrieval experiments. Finally, we present our conclusions.

\section{Related work}

Datasets have been proposed across various domains to enhance and evaluate IR systems. In this section, we present some relevant IR datasets, their contributions, and how they differ from our work.

One of the most relevant datasets for IR research is \msmarco (Microsoft Machine Reading Comprehension) \cite{msmarco}, an English-language dataset comprising 8.8 million passages (short texts) extracted from 3.2 million documents and over 530\,000 queries, each associated with at least one relevant passage. \msmarco has been automatically translated into 13 languages, including Portuguese, resulting in the creation of \mmarco \cite{mmarco}, the multilingual version of \msmarco.

The \clerc dataset is derived from the Caselaw Access Project, which contains over 1.8 million US federal case documents in English \cite{clerc, cap2024}. Adapted for information retrieval tasks, the corpus is designed to assist legal professionals by facilitating the retrieval of relevant cases and the generation of analytical texts grounded in precedents.

Within the legislative domain, the \ulyssesrfcorpus \cite{camara_dep_ir} is a relevance feedback corpus designed for information retrieval of legislative documents. It is the first publicly available corpus in Brazilian Portuguese to incorporate relevance feedback data, with annotations provided by domain experts from the Brazilian Chamber of Deputies. The corpus contains 105\,681 documents (bills) and 693 queries with user feedback. The queries represent a parliamentarian’s request for new bill proposals (or other legislative documents), with the goal of retrieving bills that address topics similar to those requested. As a result, the query format diverges from those typically used in search systems.

\quati is a Portuguese-language dataset created for information retrieval tasks \cite{bueno2024quati}. It was derived from the Portuguese subset of ClueWeb22 Category B \cite{overwijk2022clueweb22} and includes two corpora, both consisting of passages of approximately 1\,000 characters from web pages primarily written in Brazilian Portuguese: one corpus contains 1 million passages and the other has 10 million passages. Furthermore, 200 human-created queries were proposed and automatically annotated using a multi-stage pipeline involving query retrieval and evaluation by large language models (LLMs).

Finally, Silva Junior et al. introduced four Portuguese-language datasets designed for semantic textual similarity (STS) tasks focused on the legal domain \cite{datasets_legal_semantic_textual_similarity}. Two of these datasets, named \tcuvotes and \stjjudgments, consist of 371 and 7\,407 texts, respectively, extracted from the websites of TCU and the Superior Court of Justice of Brazil, without annotations. The other two datasets consist of text pairs derived from the first two, annotated for semantic similarity tasks. These datasets support the retrieval of similar documents in the legal context.

Table \ref{tab:comparison_datasets} summarizes the characteristics of these datasets. In this paper, we present the \juristcu, which provides a corpus of 16\,045 Brazilian Portuguese documents (jurisprudence from the Brazilian Federal Court of Accounts) and 150 queries with relevance judgments, suitable for evaluating search systems. The dataset with the most similar characteristics is \ulyssesrfcorpus, which, however, does not provide queries in the typical format used in search systems. Thus, we aim to bridge this gap in the literature.

\begin{table}[htbp]
\caption{Comparison of datasets.}
\renewcommand{\arraystretch}{1.1}
\centering
\begin{tabular}{llllp{5.5cm}}
    \toprule
    \textbf{Dataset}  & \textbf{Domain} & \textbf{Language} & \textbf{Corpus} & \textbf{Characteristics} \\ \midrule
    \msmarco \cite{msmarco} & General & English & 8.8 million & Passages \\
    \mmarco \cite{mmarco} & General & Multilingual & 8.8 million & Passages \\
    \clerc \cite{clerc} & Legal & English & 1.8 million & Full-text \\           
    \quati \cite{bueno2024quati} & General & Portuguese & 10 million & Passages \\
    \ulyssesrfcorpus \cite{camara_dep_ir} & Legal & Portuguese & 105\,681 & Full-text. Queries are request for new bill proposals. The goal is to find similar bills. \\
    \tcuvotes \cite{datasets_legal_semantic_textual_similarity} & Legal & Portuguese & 371 & Full-text. The goal is to find similar texts. \\
    \stjjudgments \cite{datasets_legal_semantic_textual_similarity} & Legal & Portuguese & 7\,407 & Full-text. The goal is to find similar texts. \\ \midrule
    \juristcu (ours) & Legal & Portuguese & 16\,045 & Full-text. Queries created specifically for use in search systems. \\ \bottomrule
\end{tabular}
\label{tab:comparison_datasets}
\end{table}

In addition to existing datasets, it is worth noting that LLMs have been extensively explored and adapted for legal applications. Several models have been trained or fine-tuned specifically for the legal domain, such as SaulLM-7B, SaulLM-54B, and SaulLM-141B \cite{saullm7b, saullm141b}; and BERT-like models \cite{hirs}. To evaluate the performance of LLMs in legal contexts, several benchmarks have been proposed across jurisdictions and languages, such as LawBench \cite{lawbench}, LexGLUE \cite{lexglue}, and LEXTREME \cite{lextreme}. Some professional legal exams, such as the U.S. Uniform Bar Exam \cite{gptusbarexam}, and the Brazilian Bar exam \cite{sabia3, chatgptoab} have also been used to assess the legal reasoning capabilities of LLMs. In retrieval-augmented contexts, LexRAG evaluates multi-turn legal consultation using RAG techniques \cite{lexrag}, while LegalBench-RAG focuses specifically on precision in legal text retrieval \cite{legalbenchrag}.

\section{Methodology}

The TCU "selected jurisprudence" database is available through an online search system based on BM25\footnote{\url{https://pesquisa.apps.tcu.gov.br/pesquisa/jurisprudencia-selecionada}}. This tool is widely used by various stakeholders, including public sector managers who use it to assess the legality of specific actions, and legal professionals who rely on it for client representation or legal research. In 2024, it was accessed over 1.5 million times, demonstrating its worth.

Given its importance, the online search tool should be continuously improved. This paper address this by providing qrels (query relevance judgments) for the "selected jurisprudence" database. By establishing a relevance matrix between user queries and documents, the qrels support the evaluation and enhancement of search systems. This contribution also benefits the Portuguese-language IR research community with a valuable database and annotated queries.

This section describes the design of \juristcu, an IR dataset derived from the "selected jurisprudence" database. First, we provide a brief background on this database. Then, we cover the document corpus, the queries, and the query relevance judgments.

\subsection{A brief overview of the TCU's role and its jurisprudential databases}

The \juristcu dataset was built using documents produced by TCU, which is responsible for overseeing the use of public resources in Brazil. When potential irregularities or legal issues are identified during audits, the matter is judged by a collegiate body composed of ministers. The decision is reported in an official document called an Acórdão, which is authored by a designated minister known as the rapporteur.

Each Acórdão generally includes a summary, which briefly states the decision; a report, which outlines the facts, evidence, and procedural background; and the vote, which presents the reasoning and legal interpretation adopted by the rapporteur and approved by the plenary.

These documents can be extensive, often spanning dozens of pages, and represent how the TCU interprets and applies legal norms in concrete situations. With over 500\,000 documents dating back to 1992 and made publicly available, they serve as precedents - a legal decision or form of proceeding serving as an authoritative rule or pattern in future similar cases.

These precedents collectively form the jurisprudence of the TCU. While not legally binding in the same way as in common law jurisdictions, TCU jurisprudence has normative weight and is frequently cited to support consistent interpretation of administrative law and public finance rules in Brazil.

To facilitate access to its jurisprudence, the TCU maintains a curated collection called "selected jurisprudence", which includes cases deemed to have high precedential value. This subset is the basis for the \juristcu dataset.

Each document in the "selected jurisprudence" dataset is a synthesis of a prior decision, typically derived from an Acórdão. These documents have a concise statement of the legal understanding extracted form the decision (a summary), and a chunk of text with the original decision that provides textual support for that statement (an excerpt).

\subsection{Document corpus: data collection and description of the documents}

TCU provides some of its data in CSV format on its website\footnote{\url{https://sites.tcu.gov.br/dados-abertos/}}. The \juristcu document corpus was created by downloading one of these files\footnote{\url{https://sites.tcu.gov.br/dados-abertos/jurisprudencia/arquivos/jurisprudencia-selecionada/jurisprudencia-selecionada.csv}}. While this file is currently updated weekly, data has been limited to June 2023.

Each document in this dataset represents a precedent extracted from a decision. The documents are structured into fields that include metadata about the decision (such as the rapporteur and index terms), a textual excerpt from which the jurisprudence was derived, and a summary of the decision presented as a statement. Figure \ref{fig:example_document} illustrates a sample document, while Table \ref{tab:dataset_docs} details the document structure.

\begin{figure}[!htb]
\centering
\begin{tcolorbox}[style_prompt]
\begin{verbatim}
"KEY": "JURISPRUDENCIA-SELECIONADA-1689",
"NUMACORDAO": 354.0,
"ANOACORDAO": 2016.0,
"COLEGIADO": "Plenário",
"AREA": "Licitação",
"TEMA": "Qualificação econômico-financeira",
"SUBTEMA": "Índice contábil",
"ENUNCIADO": "<b>SÚMULA TCU 289:</b> A exigência de índices contábeis...",
"EXCERTO": "<b>Fundamento Legal:</b>\\n- Decreto-Lei nº 200/1967",
"NUMSUMULA": 289.0,
"DATASESSAOFORMATADA": "24/02/2016",
"AUTORTESE": "JOSÉ MUCIO MONTEIRO",
"FUNCAOAUTORTESE": "RELATOR",
"TIPOPROCESSO": "ADMINISTRATIVO",
"TIPORECURSO": null,
"INDEXACAO": ["Exigência", "Súmula", "Justificativa"],
"INDEXADORESCONSOLIDADOS": "AREA: Licitação ; TEMA: Qualificação...",
"PARAGRAFOLC": "Processo de natureza administrativa apreciou a...",
"REFERENCIALEGAL": null,
"PUBLICACAOAPRESENTACAO": ["http://contas.tcu.gov.br/sisdoc/ObterDoc..."],
"PARADIGMATICO": "SUMULA"
\end{verbatim}
\end{tcolorbox}
\caption{Sample document.}
\label{fig:example_document}
\end{figure}

\begin{table}[htbp]
\caption{Document structure.}
\renewcommand{\arraystretch}{1.1}
\centering
{%
\begin{tabular}{p{4.5cm}p{4.6cm}p{4.5cm}}
\toprule
\textbf{Field} & \textbf{Description} & \textbf{Type} \\ \hline
KEY & Identifier Key & Text \\
NUMACORDAO & Decision number & Number \\
ANOACORDAO & Decision year & Number \\
COLEGIADO & Collegiate & Text (e.g., plenary) \\
AREA & Area indexer & Text (e.g., personnel, bidding) \\
TEMA & Topic indexer & Text (e.g., debt, fifths) \\
SUBTEMA & Subtopic indexer & Text (e.g., earnings, prohibition) \\
ENUNCIADO & Summary & Text \\
EXCERTO & Document excerpt & Text \\
NUMSUMULA & Legal summary (precedent) number & Number \\
DATASESSAOFORMATADA & Date of the judgment session & Date (DD/MM/YYYY) \\
AUTORTESE & Author of the legal thesis & Text \\
FUNCAOAUTORTESE & Role of the author & Text (e.g., rapporteur) \\
TIPOPROCESSO & Type of the process & Text (e.g., denunciation, accounts) \\
TIPORECURSO & Type of the appeal & Text (e.g.: review request) \\
INDEXACAO & Generic indexers & Text (e.g., requirement) \\
INDEXADORESCONSOLIDADOS & All Indexers & Text \\
PARAGRAFOLC & Paragraph on bidding and contracts & Text \\
REFERENCIALEGAL & Legal reference & Text \\
PUBLICACAOAPRESENTACAO & URL of the publication & URL \\
PARADIGMATICO & Paradigmatic type indexer & Text (e.g., consultation) \\
\bottomrule
\end{tabular}%
}
\vspace{0.05cm}
\label{tab:dataset_docs}
\end{table}

The corpus contains 16\,045 documents selected from the TCU's jurisprudence by domain experts from the Court. The most relevant field are ENUNCIADO and EXCERTO (hereafter referred to as SUMMARY and EXCERPT), as they convey the legal understanding derived from the decision and the textual basis for that understanding. The SUMMARY and the EXCERPT have average lengths of 306 and 4\,387 characters (47 and 660 words).

\subsection{Queries: data collection, data preparation, and description of the queries}

The TCU search system stores anonymized user data. Data from the 12 months prior to extraction of the document corpus (June 2022 to May 2023) was used to generate queries based on the most frequently used keywords and the most frequently accessed documents in the "selected jurisprudence" search system during that period. These queries were organized into three groups.

The first group of queries (G1) includes the 50 most used queries. These queries were generated by real users and show their actual interactions with the search tool. They are in keyword format, without qualifiers, punctuation, prepositions, or other non-essential elements, in line with the type of search method used (BM25). Over time, users have likely adapted to this format, recognizing that entering essential and well-chosen keywords is often enough for a successful search. On average, the queries in this group have 3.5 words.

The second and third groups of queries (G2 and G3), each consisting of 50 queries, were created by considering the 50 most accessed documents during that period. For each document, the SUMMARY was used to prompt an LLM (ChatGPT 4) to generate five questions answerable by that summary (Figure \ref{fig:prompt_g2_g3} in Appendix \ref{sec:prompts_pt} shows the prompt used\footnote{All the original prompts in Portuguese are presented in Appendix \ref{sec:prompts_pt}. Appendix \ref{sec:prompts_en} provides their English translations.}). These five questions were manually refined into a single, representative question that accurately reflects the SUMMARY. Based on this refined question, two groups of queries were created: a keyword-based version (G2), which contains only essential words (similar to G1), and a full-text version (G3), which includes the complete question. The queries in groups G2 and G3 average 6.5 and 16.5 words in length, respectively.

\begin{figure}[!htbp]
    \centering
    \includegraphics[width=0.9\linewidth]{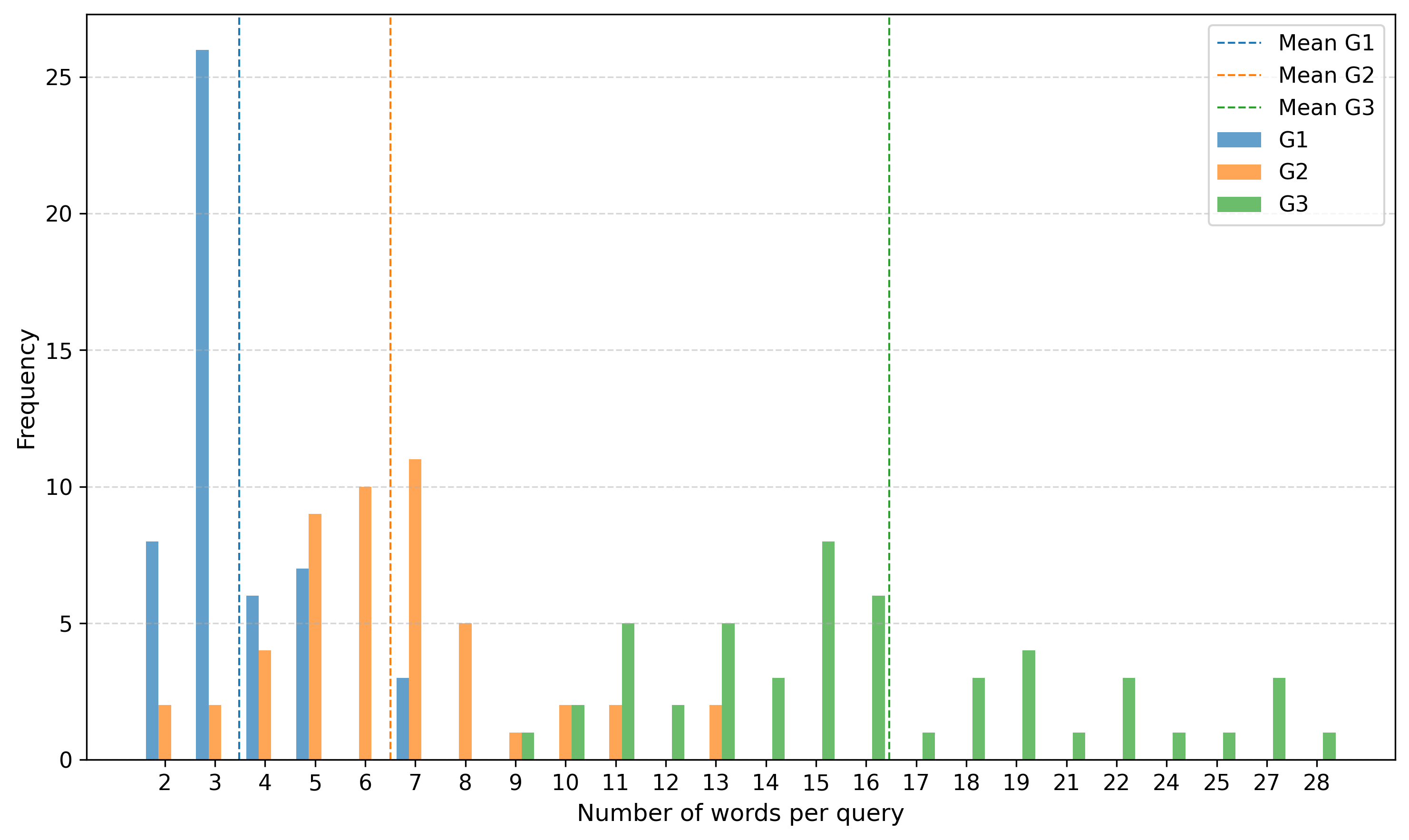}
    \caption{Histogram of the number of words per query for the query groups G1, G2, and G3.}
\label{fig:histogram}
\end{figure}

This process resulted in three query groups, whose characteristics are summarized in Table \ref{tab:dataset_queries}. Figure \ref{fig:histogram} shows the histogram showing the number of words per query for each group. Tables \ref{tab:G1}, \ref{tab:G2}, and \ref{tab:G3} in Appendix \ref{sec:anexo_queries} list all queries for groups G1, G2, and G3.

\begin{table}[htbp]
\caption{Characteristics of the query groups.}
\centering
{%
\begin{tabular}{lllc}
\toprule
\textbf{Query group} & \textbf{Description} & \textbf{Format} & \textbf{Average words per query} \\ \midrule
G1 & 50 most used queries & Keyword-based & 3.5 \\
G2 & 50 synthetic queries & Keyword-based & 6.5 \\
G3 & 50 synthetic queries & Question-based & 16.5 \\ \bottomrule
\end{tabular}%
}
\vspace{0.05cm}
\label{tab:dataset_queries}
\end{table}

\subsection{Relevance judgment annotation: qrels generation}

Ideally, with unlimited resources, query annotation requires clear relevance criteria and a multidisciplinary team of experts to independently evaluate, score, and potentially reach a consensus on all documents per query. This scenario is usually unfeasible due to the high cost and time required for annotation \cite{llmassistedrelevanceassessmentsask}. Consequently, alternative methods are often used, such as annotating only a subset of documents using search tools pre-selection of potentially relevant items \cite{construction_large_test_collections}, using crowd workers instead of experts \cite{llm_can_predict_search_preferences}, and, more recently, using LLM-based automatic evaluations \cite{llmassistedrelevanceassessmentsask}.

Each approach has trade-offs. Annotating only a subset of documents increases the risk of missing relevant ones. Additionally, using crowd workers instead of domain experts may lead to misinterpretations of user needs, especially in specialized corpora such as legal texts. Furthermore, the use of LLMs for evaluation also poses the risk of hallucinations or bias.

Considering these trade-offs, we adopted a two-step process: an initial automated evaluation using LLMs, followed by a manual review by a domain expert to correct the judgments produced by the automated process. This approach aligns with the use of multi-stage retrievers, which are commonly employed in IR due to their strong performance \cite{multistagesystems} and  accelerates the creation of qrels while still preserving a degree of human oversight

The annotation process starts with a two-stage retriever to select potentially relevant documents for each query. The first stage selects a subset of candidate documents, and the second stage reranks them. Although classical lexical retrievers are usually employed in the first stage, the use of semantic retrievers in this stage has also been discussed \cite{multistagesystems}. Therefore, in the first stage we used both retrievers—a lexical one (BM25) and a semantic one (rufimelo/Legal-BERTimbau-sts-large-ma-v3 BERT-based model\footnote{\url{https://huggingface.co/rufimelo/Legal-BERTimbau-sts-large-ma-v3}}). Considering all queries, the overlap between the top 300 documents returned by both retrievers ranged from 1.3\% to 44.6\%, averaging 15.89\% across the 150 queries. The top 300 documents returned by each retriever are combined and then sent to the second stage, where the 600 documents are reranked using a T5-based model (unicamp-dl/mt5-3B-mmarco-en-pt\footnote{\url{https://huggingface.co/unicamp-dl/mt5-3B-mmarco-en-pt}}). The top 10 ranked results are selected for annotation, along with 5 randomly chosen documents from the top 1\,000 documents retrieved by BM25.

The purpose of selecting 5 random documents from the BM25 output is to provide documents with different levels of relevance for annotation, as these documents are highly likely to be irrelevant. This approach is commonly used when training ranking models (e.g., \cite{splade}), where both relevant and non-relevant examples are needed. Non-relevant documents can be selected randomly from the corpus (random negatives) or from retriever outputs (hard negatives) \cite{hard_negatives_1, hard_negatives_2}. Hard negatives, like those from BM25, share some terms with the query but are not truly relevant, making them more useful for evaluating retrieval methods.

These 15 documents are sent to an LLM for automatic annotation, where relevance scores are assigned on a scale from 0 (irrelevant) to 3 (highly relevant). The evaluation used the GPT-4 model with the prompt in Figure \ref{fig:prompt_score_relevancia} in Appendix \ref{sec:prompts_pt}. As a result, each query is expected to have 15 annotated documents with varying degrees of relevance.

While some researchers suggest that LLMs could potentially replace human evaluation, the extent of their universal applicability remains a subject of ongoing debate \cite{llm_cant_replace_human_relevance_assessment}. To address this, a domain expert reviewed the automatically generated relevance judgments to correct them if necessary, resulting in the final qrels. In the case of \juristcu, the expert agreed with the automatic evaluation. We believe two factors contributed to this agreement. First, LLMs used in legal contexts have often produced results that outperform those of human lawyers \cite{legalbenchpt}. Second, expert validation of pre-generated results can introduce confirmation bias. Unless the output is highly inconsistent (e.g., a clearly irrelevant document marked as highly relevant), experts tend to agree with the judgments they are reviewing.

\begin{figure}[!htb]
    \centering
    \includegraphics[width=1\linewidth]{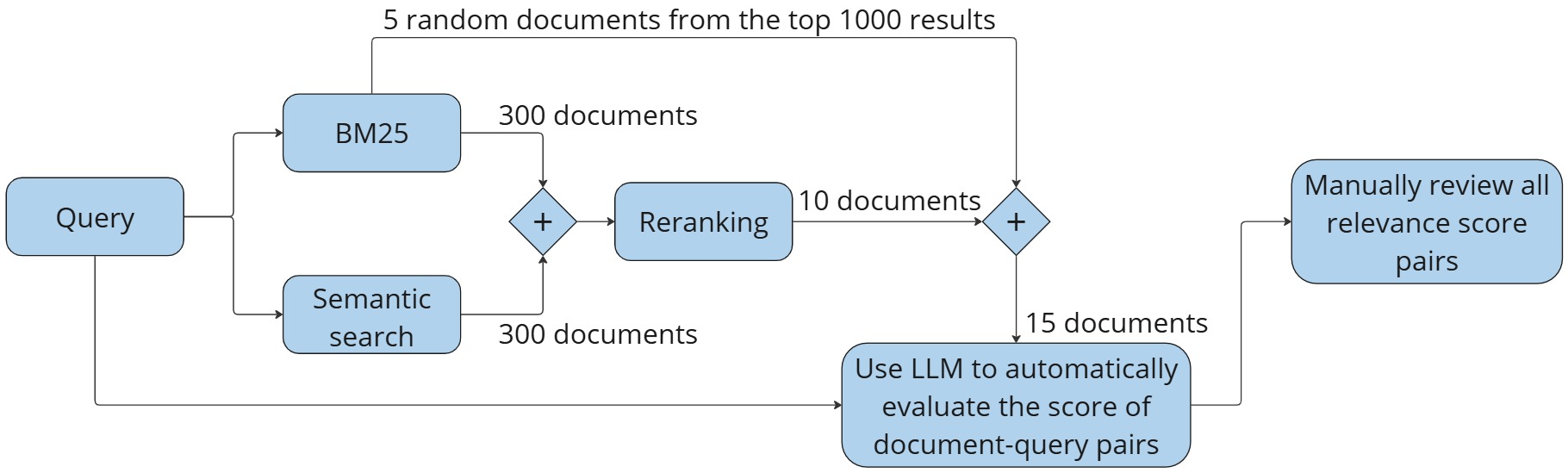}
    \caption{Process to generate the qrels.}
\label{fig:pipeline_docs_selection}
\end{figure}

Figure \ref{fig:pipeline_docs_selection} summarizes the annotation pipeline.

\section{Information retrieval experiments using the \juristcu dataset}

This section details the IR experiments conducted with the \juristcu dataset. We describe the experimental setup, present the analysis, and discuss the results, which serve as a benchmark for the dataset and contribute to the improvement of the TCU's search system.

\subsection{Experimental setup}

The experimental setup evaluates different retrievers using the \juristcu dataset. Since the TCU's jurisprudence search system relies on a BM25-based lexical search, we used a standard BM25 implementation as the baseline rather than directly assessing the TCU system, providing a well-established and widely recognized benchmark for comparison.

Traditional information retrieval models like BM25 are lexical, relying on exact word matching between the query and the document, without considering their meaning or context. A key limitation of lexical approaches is vocabulary mismatch, which occurs when a user's query contains keywords that are semantically equivalent or similar to those in a relevant document but differ lexically \cite{wang2024utilizing}. As a result, traditional methods may assign a lower relevance score than expected if semantics were considered, potentially preventing relevant documents from being retrieved.

There are three general approaches to address this issue \cite{lin2022pretrained}:

\begin{itemize}
    \item Expand queries: The original query is augmented  with additional terms to improve retrieval effectiveness. This can be done using synonym dictionaries or more sophisticated techniques.
    
    \item Expand documents: Similar to query expansion, this technique enriches documents with new textual information prior to indexing \cite{nogueira2019doc2query}.
    
    \item Go beyond exact term matching: This method converts the representation of queries and documents from a lexical to a semantic space, enabling retrieval based on meaning rather than exact word matching \cite{lin2022pretrained}.
\end{itemize}

The first two approaches, query and document expansions, offer the practical advantage of leveraging existing search systems, which are predominantly lexical, as they modify only the data and not the underlying infrastructure, such as code or textual databases. Traditionally, query expansion has been more widely used because document expansion is harder to test, as it requires reindexing the entire database. However, with the advent of LLMs, document expansion techniques have demonstrated superior performance compared to query expansion \cite{nogueira2019document}, so we did not pursue experimentation with the latter.

Thus, we compare the standard BM25 baseline with two main retrieval approaches: lexical retrieval with document expansion and semantic retrieval. Our experiments focus solely on the SUMMARY and EXCERPT fields, as they provide the most relevant information for retrieval tasks.

\subsubsection{Lexical retrieval with document expansion}

In the lexical retrieval experiments, we indexed the SUMMARY and EXCERPT fields using BM25. We also applied document expansion techniques that augments the content of the documents with supplementary text. In these experiments, the expansion used the SUMMARY as input, and the expanded text is indexed with the original document. We evaluated two document expansion methods:

\begin{itemize}
    \item docT5query \cite{nogueira2019doc2query}: This method uses a T5-based model to generate synthetic queries that could be used to retrieve a given document. In this experiment, we used a model trained in Portuguese (doc2query/msmarco-portuguese-mt5-base-v1\footnote{\url{https://huggingface.co/doc2query/msmarco-portuguese-mt5-base-v1}}) to generate five synthetic queries from the SUMMARY, which were then appended to the document;

    \item Synonym expansion with LLMs: This method uses LLMs to generate synonyms and related terms for relevant words of the SUMMARY. We considered three LLMs (GPT-3.5, GPT-4, and Llama 3-70B) and used the prompt shown in Figure \ref{fig:prompt_synonym_expansion} in Appendix \ref{sec:prompts_pt}, which requests two or three synonyms for the five most relevant words in the SUMMARY. The text generated was appended to the document.
\end{itemize}

We also evaluated combinations of both expansion methods. Both methods enhance the lexical representation of documents, increasing the likelihood of retrieving relevant documents based on the contribution of important words.

\subsubsection{Semantic retrieval}

For semantic retrieval, we used LLMs to generate dense vector representations of the SUMMARY, enabling the retriever to identify documents semantically similar to the query. We tested two groups of embedding models:

\begin{itemize}
    \item BERT-based models: We evaluated one multilingual model and four models fine-tuned on Portuguese text. Among the models focused on Portuguese, one was fine-tuned on general texts and the others on texts in the legal domain, including TCU legal documents.
    
    \item OpenAI models: We also experimented with embeddings generated by OpenAI models, known for their strong generalization capabilities across various domains.
\end{itemize}

All embeddings were generated from the SUMMARY field. For the BERT-based models, since the input token size is limited, the input was truncated when necessary.

\subsubsection{List of experiments}
\label{sec:list_experiments}

We conducted 14 experiments: one baseline (BM25), seven with document expansion, and six using semantic representations of texts. The lexical experiments use BM25 with $k_1 = 1.2$ and $b = 0.75$\footnote{\url{https://www.elastic.co/pt/blog/practical-bm25-part-3-considerations-for-picking-b-and-k1-in-elasticsearch}}. Table \ref{tab:list_experiments} details all of them.

\begin{table}[htbp]
\caption{Description of the experiments.}
\centering
{%
\renewcommand{\arraystretch}{1.3}
\begin{tabular}{@{}lp{11cm}@{}}
\toprule
\begin{tabular}[c]{@{}c@{}}\textbf{Experiment}\end{tabular} & \begin{tabular}[c]{@{}c@{}}\textbf{Description}\end{tabular}\\ \midrule
BM25 (baseline) & This is the baseline, a simple BM25 retriever. In the analyses, unless stated otherwise, all comparisons are made with the baseline. \\ \midrule
BM25.dT5q & BM25. Documents expanded with 5 queries generated by the docT5query method. \\ 
BM25.Syn(GPT3.5) & BM25. Documents expanded with synonyms generated by the GPT-3.5 model. \\ 
BM25.Syn(GPT4o) & BM25. Documents expanded with synonyms generated by the GPT-4o model. \\ 
BM25.Syn(Llama3) & BM25. Documents expanded with synonyms generated by the Llama-3-70B model. \\ 
BM25.dT5q.Syn(GPT35) & BM25. Documents expanded with the docT5query method and synonyms generated by the GPT-3.5 model. \\ 
BM25.dT5q.Syn(GPT4o) & BM25. Documents expanded with the docT5query method and synonyms generated by the GPT-4o model. \\ 
BM25.dT5q.Syn(Llama3) & BM25. Documents expanded with the docT5query method and synonyms generated by the Llama-3-70B model. \\ \midrule
BERT.pt.TCU & 768-dimensional embeddings from the Luciano/bert-base-portuguese-cased-finetuned-tcu-acordaos\tablefootnote{\url{https://huggingface.co/Luciano/bert-base-portuguese-cased-finetuned-tcu-acordaos}} model, which was fine-tuned from a Portuguese BERT-base model (BERTimbau) using TCU documents. \\ 
BERT.pt.large & 1\,024-dimensional embeddings from the neuralmind/bert-large-portuguese-cased\tablefootnote{\url{https://huggingface.co/neuralmind/bert-large-portuguese-cased}} model (aka. BERTimbau-large) \cite{souza2020bertimbau}, a BERT model fine-tuned on Portuguese texts. \\ 
BERT.pt.large.legal & 1\,024-dimensional embeddings from stjiris/bert-large-portuguese-cased-legal-mlm-sts-v1\tablefootnote{\url{https://huggingface.co/stjiris/bert-large-portuguese-cased-legal-mlm-sts-v1.0}} model \cite{MeloSemantic} (aka. Legal BERTimbau-large), a BERT model fine-tuned using legal text of the Supreme Court of Justice of Portugal. \\ 
BERT.ml & 768-dimensional embeddings from the multilingual paraphrase-multilingual-mpnet-base-v2\tablefootnote{\url{https://huggingface.co/sentence-transformers/paraphrase-multilingual-mpnet-base-v2}} BERT model \cite{bertmultilingua}. \\ 
OpenAI.small & 1\,536-dimensional embeddings generated using the general-purpose and multilingual text-embedding-3-small model by OpenAI. \\ 
OpenAI.large & 3\,072-dimensional embeddings generated using the general-purpose and multilingual text-embedding-3-large model by OpenAI. \\ \bottomrule
\end{tabular}%
}
\vspace{0.05cm}
\label{tab:list_experiments}
\end{table}

\subsection{Experiments results}

The 150 queries from the \juristcu dataset were used to evaluate the experiments listed in Table \ref{tab:list_experiments}. Given the distinct characteristics of the query groups, the results are presented separately for each of the three groups (G1, G2, and G3). The following metrics were calculated: precision (P@10), recall (R@10), mean reciprocal rank (MRR@10), and normalized discounted cumulative gain (nDCG@10). These metrics assess whether relevant documents are retrieved and whether their ranking is appropriate. To calculate these metrics, documents with a score different from 0 were considered relevant to some degree.

\FloatBarrier

\subsubsection{Query group 1 - G1}

\begin{figure}[!htbp]
    \centering
    \includegraphics[width=1\linewidth]{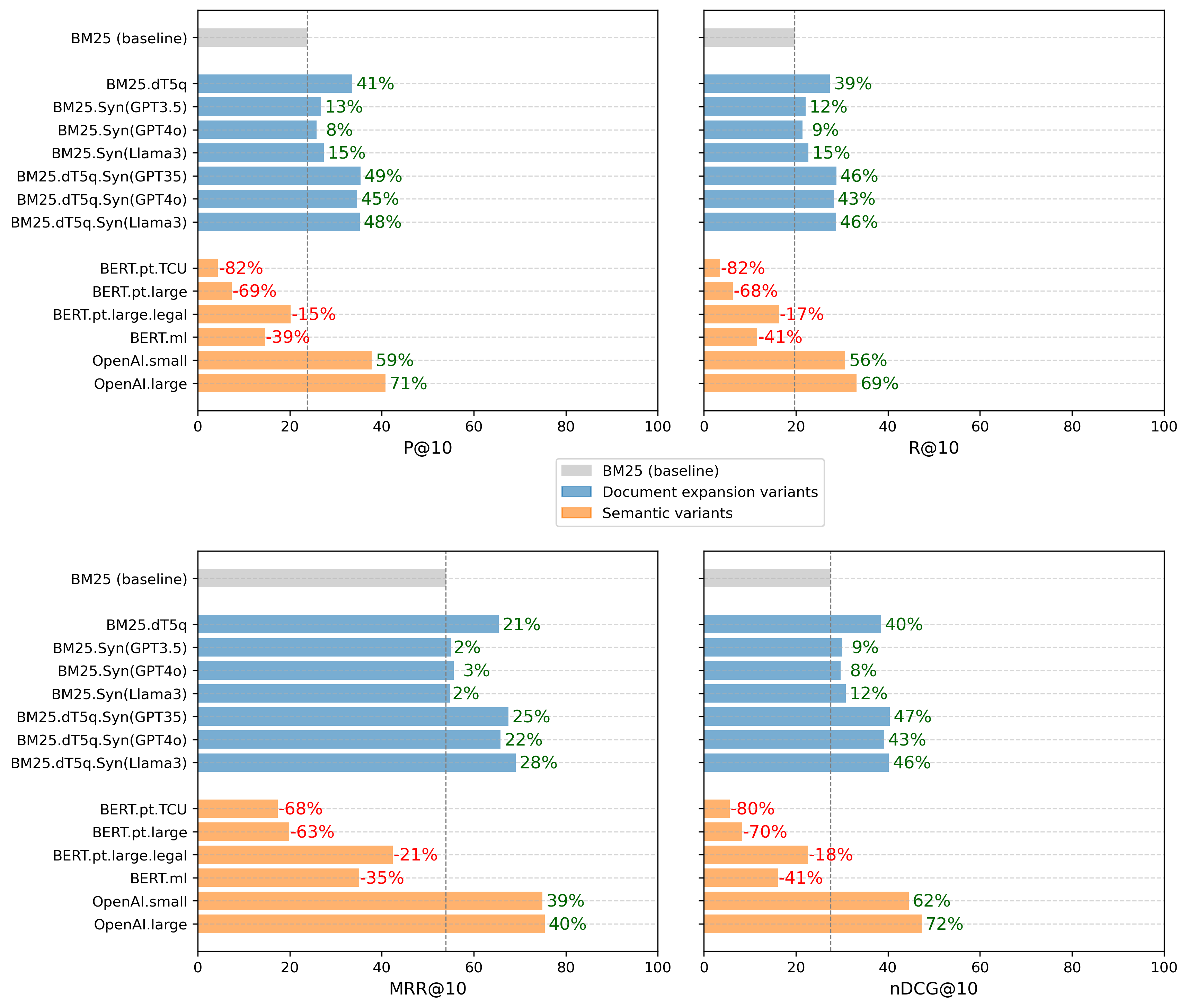}
    \caption{Performance metrics for query group G1. The percentages on each bar show how much the metric improved/worsened compared to the baseline.}
\label{fig:results_at_10_g_1}
\end{figure}

Group G1 consists of the 50 most frequent queries submitted by users of the TCU's search system. They represent real-world search behavior, are keyword-based and average 3.5 words in length. Table \ref{tab:results_at_10_g_1} and Figure \ref{fig:results_at_10_g_1} present the performance metrics for the various experiments evaluated with this query group.

\begin{table}[htbp]
\caption{Performance metrics for query group G1.}
\centering
{%
\begin{tabular}{@{}lrrrr@{}}
\toprule
	& \begin{tabular}[c]{@{}c@{}}\textbf{P@10}\end{tabular} 
	& \begin{tabular}[c]{@{}c@{}}\textbf{R@10}\end{tabular}
        & \begin{tabular}[c]{@{}c@{}}\textbf{MRR@10}\end{tabular}
	& \begin{tabular}[c]{@{}c@{}}\textbf{nDCG@10}\end{tabular}\\ \midrule
BM25 (baseline)       & 23.8          & 19.7          & 53.9          & 27.5 \\ \midrule
BM25.dT5q             & 33.6          & 27.4          & 65.4          & 38.5 \\
BM25.Syn(GPT3.5)      & 26.8          & 22.1          & 55.1          & 30.1 \\
BM25.Syn(GPT4o)       & 25.8          & 21.4          & 55.6          & 29.7 \\
BM25.Syn(Llama3)      & 27.4          & 22.7          & 54.8          & 30.8 \\
BM25.dT5q.Syn(GPT35)  & 35.4          & 28.8          & 67.5          & 40.4 \\
BM25.dT5q.Syn(GPT4o)  & 34.6          & 28.2	      & 65.8          & 39.2 \\
BM25.dT5q.Syn(Llama3) & 35.2          & 28.7          & 69.1          & 40.2 \\ \midrule
BERT.pt.TCU           & 4.4           & 3.5           & 17.4          & 5.6 \\
BERT.pt.large         & 7.4           & 6.3           & 19.9          & 8.3 \\
BERT.pt.large.legal   & 20.2          & 16.3          & 42.4          & 22.6 \\
BERT.ml               & 14.6          & 11.6          & 35.1          & 16.1 \\
OpenAI.small          & 37.8          & 30.7          & 74.9          & 44.5 \\
OpenAI.large          & \textbf{40.8} & \textbf{33.2} & \textbf{75.4} & \textbf{47.3} \\\bottomrule
\end{tabular}%
}
\vspace{0.05cm}
\label{tab:results_at_10_g_1}
\end{table}

All document expansion methods significantly improved the metrics compared to the baseline (BM25). They retrieved more relevant documents (P@10 and R@10), improved the ranking of the first relevant document (MRR@10), and enhanced the overall ranking quality of the top 10 retrieved documents (nDCG@10).

Within the lexical domain, the docT5query method outperformed document expansion prompting an LLM. Compared to the baseline, docT5query improved P@10 from 23.8 to 33.6 (41\% increase). On the other hand, the best synonym expansion method increased precision to 27.4 (15\% increase compared to the baseline). Similar results were also observed for R@10 and nDCG@10. There were also improvements in MRR@10, but to a lesser extent (21\% improvement with docT5query).

The best results using document expansion were achieved by combining both techniques. For instance, the highest P@10 was 35.4 (49\% improvement over the baseline). Among the LLMs tested, GPT-4o generally performed worse than GPT-3.5 or Llama 3-70B. For this particular task, using a larger and more computationally expensive LLM did not improved retrieval quality. Additionally, no consistent performance difference was observed between GPT-3.5 and Llama 3-70B.

Among the semantic retrieval models, the BERT-base models performed significantly worse than the BM25 baseline. Comparing the three experiments with models fine-tuned in Portuguese (experiments with the prefix BERT.pt), we observe that the worst result was obtained with the BERT.pt.TCU experiment, followed by BERT.pt.large and BERT.pt.large.legal. Considering only the model sizes, this ranking is not surprising, as the model used in the first experiment is the smallest of the three (768-dimensional embeddings). Additionally, between the last two models (1\,024-dimensional embeddings), a better result was expected from the BERT.pt.large.legal experiment, as it was fine-tuned on legal domain texts. However, it is surprising that the performance of the BERT.pt.TCU and BERT.pt.large experiments was worse than that of the BERT.ml experiment, which used a multilingual model with 768-dimensional embeddings.

No BERT-based model outperformed the baseline. There are some possible explanations for this. Since BERT's context window is relatively small, and the SUMMARY of a few documents was truncated to fit this limit, it is possible that some information was lost. Additionally, group G1 consists of short search expressions (an average of 3.5 words per query), which poses a challenge for some semantic models, as they tend to perform better in larger contexts.

On the other hand, the OpenAI models substantially outperformed both the BM25 baseline and all document expansion methods. Moreover, the larger model (3\,072 dimensions) produced better results than the smaller one (1\,536 dimensions). P@10, R@10, and nDCG@10 increased by approximately 70\% compared to the baseline, while MRR@10, which indicates the ranking of the first relevant document retrieved, improved by 40\%. Notably, the OpenAI models effectively represented the semantics, even for short queries.

An important consideration is that the way the qrels were generated, through a pre-selection of documents using BM25 and a BERT-based model, could introduce bias towards these two models, favoring documents retrieved by them. However, the best results in Table \ref{tab:results_at_10_g_1} do not come from these models; in fact, they performed poorly. This may suggest that the annotation process of the qrels, particularly the manual curation, at least partially corrected this bias introduced during the pre-selection process.

\subsubsection{Query group 2 - G2}

\begin{figure}[!htbp]
    \centering
    \includegraphics[width=1\linewidth]{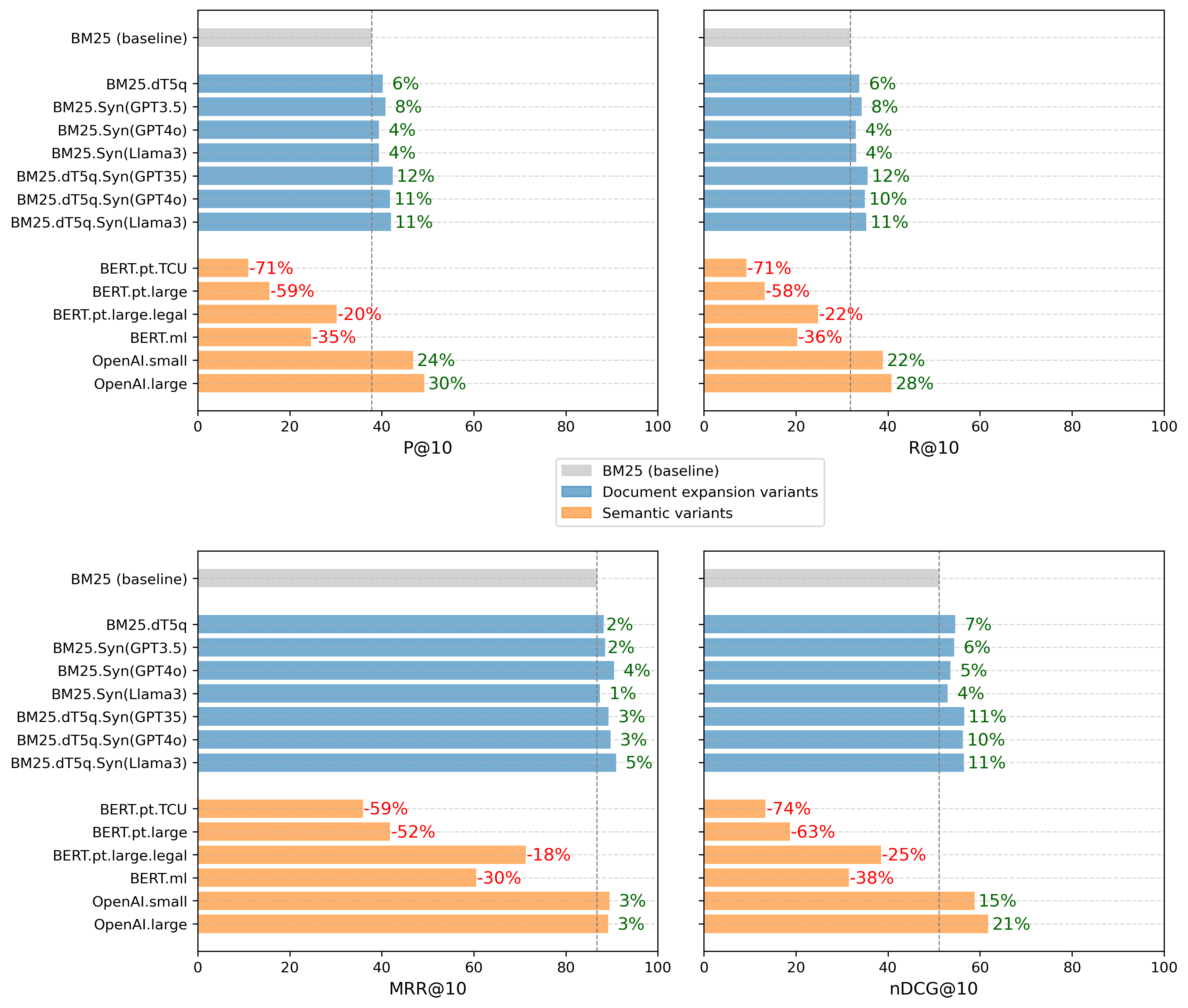}
    \caption{Performance metrics for query group G2. The percentages on each bar show how much the metric improved/worsened compared to the baseline.}
\label{fig:results_at_10_g_2}
\end{figure}

Group G2 was generated from the summaries of the 50 most accessed documents in the \juristcu dataset. Each query contains an average of 6.5 relevant keywords extracted from these documents. Table \ref{tab:results_at_10_g_2} and Figure \ref{fig:results_at_10_g_2} present the metrics for this query group.

The first noticeable aspect is the increase in baseline metrics relative to group G1. Although the query groups are not comparable, this was expected, as group G1 consists of real user queries, whereas group G2 contains synthetic queries derived from document summaries. The nature of BM25 scoring suggests that using more keywords present in the target documents will inherently lead to better retrieval performance. Consequently, in this group, BM25 serves as a more challenging benchmark compared to G1.

Compared to the BM25 baseline, document expansion techniques improved performance across all metrics. In this group, the docT5query and the synonym extraction via LLM prompting showed comparable performance. However, as in group G1, combining both methods proved to be more effective than using either one alone. Consistently with the results in group G1, GPT-4o's generally performed worse than GPT-3.5 or Llama 3-70B, and there is no clear evidence indicating the advantage of one over the latter two models.

Although all metrics improved in the lexical experiments, the increases in group G2 were substantially smaller than those in group G1. For example, the best P@10 and nDCG@10 achieved were just over 10\% higher than the baseline. This limited improvement was expected due to the high alignment of keywords present in the queries with the documents, which favors BM25-based retrievers and makes the baseline a more difficult target to surpass.

\begin{table}[!htbp]
\caption{Performance metrics for query group G2.}
\centering
{%
\begin{tabular}{@{}lrrrr@{}}
\toprule
	& \begin{tabular}[c]{@{}c@{}}\textbf{P@10}\end{tabular} 
	& \begin{tabular}[c]{@{}c@{}}\textbf{R@10}\end{tabular}
        & \begin{tabular}[c]{@{}c@{}}\textbf{MRR@10}\end{tabular}
	& \begin{tabular}[c]{@{}c@{}}\textbf{nDCG@10}\end{tabular}\\ \midrule
BM25 (baseline)       & 37.8          & 31.8          & 86.7          & 51.1 \\ \midrule
BM25.dT5q             & 40.2          & 33.8          & 88.2          & 54.6 \\
BM25.Syn(GPT3.5)      & 40.8          & 34.3          & 88.5          & 54.4 \\
BM25.Syn(GPT4o)       & 39.4          & 33.0          & 90.5          & 53.6 \\
BM25.Syn(Llama3)      & 39.4          & 33.1          & 87.4          & 53.0 \\
BM25.dT5q.Syn(GPT35)  & 42.4          & 35.6          & 89.3          & 56.6 \\
BM25.dT5q.Syn(GPT4o)  & 41.8          & 35.0          & 89.7          & 56.3 \\
BM25.dT5q.Syn(Llama3) & 42.0          & 35.3          & \textbf{90.9} & 56.5 \\ \midrule
BERT.pt.TCU           & 11.0          & 9.2           & 35.9          & 13.4 \\
BERT.pt.large         & 15.6          & 13.2          & 41.8          & 18.7 \\
BERT.pt.large.legal   & 30.2          & 24.8          & 71.3          & 38.5 \\
BERT.ml               & 24.6          & 20.3          & 60.5          & 31.5 \\
OpenAI.small          & 46.8          & 38.9          & 89.5          & 58.8 \\
OpenAI.large          & \textbf{49.2} & \textbf{40.8} & 89.2          & \textbf{61.8} \\\bottomrule
\end{tabular}%
}
\vspace{0.05cm}
\label{tab:results_at_10_g_2}
\end{table}

Regarding the semantic retrieval models, we observe the same trend observed in group G1. The BERT-based models continued to underperform compared to the baseline. However, the embeddings generated by the OpenAI models produced highly satisfactory results and, except for MRR@10, outperformed all other models across the remaining metrics. In this query group, as in group G1, the OpenAI model with the larger dimension produced better results. Even when faced with a BM25 benefiting from the manner in which the synthetic queries were generated, the OpenAI.large experiment significantly improved the metrics (increases of 30\%, 28\%, and 21\% for P@10, R@10, and nDCG@10, respectively).

\FloatBarrier

\subsubsection{Query group 3 - G3}

\begin{figure}[!htbp]
    \centering
    \includegraphics[width=1\linewidth]{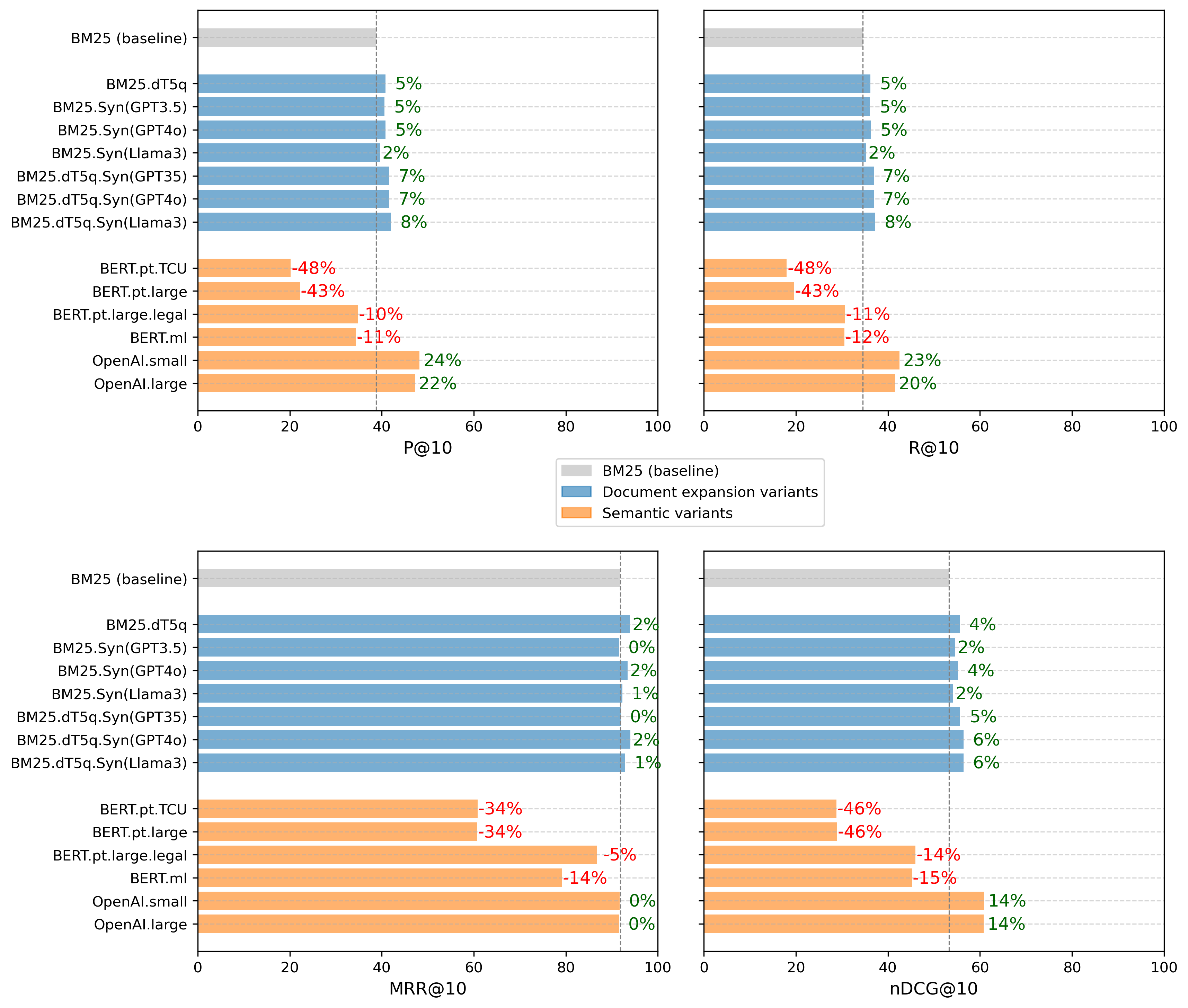}
    \caption{Performance metrics for query group G3. The percentages on each bar show how much the metric improved/worsened compared to the baseline.}
\label{fig:results_at_10_g_3}
\end{figure}

Similarly to group G2, group G3 was generated based on the 50 most accessed documents in the TCU search system. However, in this case, the queries are formulated as questions that can be answered by the SUMMARY. They have an average length of 16.5 words per query, making them the longest among all groups. Table \ref{tab:results_at_10_g_3} and Figure \ref{fig:results_at_10_g_3} present the metric results for this query group.

\begin{table}[htbp]
\caption{Performance metrics for query group G3.}
\centering
{%
\begin{tabular}{@{}lrrrr@{}}
\toprule
	& \begin{tabular}[c]{@{}c@{}}\textbf{P@10}\end{tabular} 
	& \begin{tabular}[c]{@{}c@{}}\textbf{R@10}\end{tabular}
        & \begin{tabular}[c]{@{}c@{}}\textbf{MRR@10}\end{tabular}
	& \begin{tabular}[c]{@{}c@{}}\textbf{nDCG@10}\end{tabular}\\ \midrule
BM25 (baseline)       & 38.8          & 34.5          & 91.8          & 53.3 \\ \midrule
BM25.dT5q             & 40.8          & 36.2          & 93.9 & 55.6 \\
BM25.Syn(GPT3.5)      & 40.6          & 36.1          & 91.5          & 54.6 \\
BM25.Syn(GPT4o)       & 40.8          & 36.3          & 93.4          & 55.2 \\
BM25.Syn(Llama3)      & 39.6          & 35.2          & 92.3          & 54.1 \\
BM25.dT5q.Syn(GPT35)  & 41.6          & 36.9          & 91.9          & 55.7 \\
BM25.dT5q.Syn(GPT4o)  & 41.6          & 36.9          & \textbf{94.0} & 56.4 \\
BM25.dT5q.Syn(Llama3) & 42.0          & 37.2          & 92.9          & 56.4 \\ \midrule
BERT.pt.TCU           & 20.2          & 18.0          & 60.8          & 28.8 \\
BERT.pt.large         & 22.2          & 19.6          & 60.7          & 28.9 \\
BERT.pt.large.legal   & 34.8          & 30.7          & 86.8          & 46.0 \\
BERT.ml               & 34.4          & 30.5          & 79.2          & 45.2 \\
OpenAI.small          & \textbf{48.2} & \textbf{42.5} & 91.7          &\textbf{60.9} \\
OpenAI.large          & 47.2          & 41.5          & 91.5          & 60.8 \\\bottomrule
\end{tabular}%
}
\vspace{0.05cm}
\label{tab:results_at_10_g_3}
\end{table}

In this group, the queries are longer and use vocabulary extracted directly from the documents. In this scenario, BM25 is highly effective, as it performs well when most of the words in longer queries match those in relevant documents. Thus, it becomes an even more challenging benchmark to surpass.

Nevertheless, even in this scenario, almost all experiments with document expansion techniques surpassed the baseline. As in group G2, the docT5query and the synonym extraction via LLM prompting method performed similarly. Once again, combining both methods proved to be more effective than using either one alone. However, the three LLMs tested for synonym extraction (GPT-3.5, GPT-4o, and Llama 3-70B) produced very similar results. The improvements were more modest than in the other query groups: approximately 8\% for P@10 and R@10, and 6\% for nDCG@10.

In semantic retrieval, the BERT-based models continued to perform worse than the BM25 baseline. In this query group, except for MRR@10, the multilingual model (BERT.ml experiment) performed similarly to the best-performing Portuguese model (BERT.pt.large.legal experiment). The OpenAI models achieved the best overall performance. However, the smaller OpenAI model slightly outperformed the larger one. The OpenAI.small experiment increased P@10 and R@10 by about 24\% and nDCG@10 by approximately 14\%.

The best lexical and semantic models did not significantly improve MRR@10, and for the semantic models, the score even slightly decreased. This was expected, as the queries are well aligned with the document to be retrieved, leading the baseline to almost always return a relevant document in the first position.

\FloatBarrier

\section{Conclusion}

This paper introduced \juristcu, a Brazilian Portuguese dataset designed for legal information retrieval (LIR), featuring 16\,045 jurisprudential documents from the Brazilian Federal Court of Accounts and 150 queries with relevance judgments. By addressing the absence of Portuguese LIR datasets with query created and annotated specifically for use in search systems, \juristcu enables benchmarking of retrieval systems in a specialized legal context in Portuguese.

In addition to the dataset, we conducted 14 experiments using both lexical and semantic search. For lexical search, we tested document expansion using the docT5query method and synonym-based document expansion by prompting LLMs (GPT-3.5, GPT-4o, and Llama 3-70B). In the semantic search experiments, we generated embeddings using both Portuguese and multilingual BERT-based models, as well as proprietary models from OpenAI.

The document expansion techniques, whether using docT5query or synonym extraction via LLM prompting, improved the performance metrics over the BM25 baseline. Combining both techniques yielded the best results in the lexical domain. Among the three LLMs used for synonym extraction, usually GPT-3.5 and Llama 3-70B produced the best results, with no clear advantage of one over the other.

In the lexical experiments, the most significant improvements occurred when evaluating short keyword-based queries, with over 45\% improvements for P@10, R@10, and nDCG@10. There were also gains when the queries are well-formed, question-based, and highly aligned with the documents; however, these improvements were substantially smaller (e.g., up to 8\% in P@10 and R@10, and up to 6\% in nDCG@10).

In the semantic domain, the BERT-based models underperformed the baseline. The poor performance of the BERT-based models highlights that their use for extracting semantic text embeddings in this specific domain is not recommended, which may indicate the need for fine-tuning on this dataset to achieve satisfactory results.

However, all OpenAI embeddings models consistently achieved the best metrics in the dataset, even when compared to the top-performing lexical experiments. The highest improvements over the baseline also occurred when evaluating short keyword-based queries, with increases of approximately 70\% in P@10, R@10, and nDCG@10 metrics. The superiority of the OpenAI models, even with short keyword-based queries, suggests that dense embeddings capture semantic relationships in this domain, surpassing the dependence on lexical terms.

This study has some limitations. The \juristcu dataset focuses on TCU jurisprudence, and therefore, its use to infer the performance of retrievers may not generalize to other domains (e.g., criminal law). However, similar to other datasets, it can be used to study the impact of new embedding models or search techniques.

This paper has also practical implications. Besides offering a dataset for the Portuguese-language IR research community, both the dataset and the experimental results contribute to enhance a search system that is highly relevant to Brazilian citizens, with documents accessed over 1.5 million times in 2024.

For future work, it is essential for the Portuguese-language IR community to have a diverse collection of datasets available, similar to the English-language IR community. Therefore, we plan to release new annotated datasets to support further research and development in this field. Another possibility is the manual annotation of the \juristcu dataset, which would enable progress in studies comparing manual and automatic relevance judgments in Portuguese.

\bibliography{main.bib}

\appendix

\section{Original prompts in Portuguese}
\label{sec:prompts_pt}

This appendix presents all the prompts used in this article, in Portuguese. Appendix \ref{sec:prompts_en} shows their translated version into English.

\begin{figure}[!htb]
\centering
\begin{tcolorbox}[style_prompt]
\textbf{[User]}

Elabore até 5 perguntas curtas e diretas que possam ser respondidas a partir do enunciado a seguir:\\

\{ENUNCIADO\}
\end{tcolorbox}
\caption{Prompt used to generate queries of groups G2 and G3.}
\label{fig:prompt_g2_g3}
\end{figure}

\begin{figure}[!htb]
\begin{tcolorbox}[style_prompt]
\textbf{[System]}\\
Você é um especialista na jurisprudência do Tribunal de Contas da União com o objetivo de  avaliar se um enunciado de jurisprudência responde a uma pergunta.\\
Deve retornar um valor de escore de 0 a 3, sendo: \\
0 - irrelevante - o enunciado não responde a pergunta; \\
1 - relacionado - o enunciado apenas está no tópico da pergunta; \\
2 - relevante - o enunciado responde parcialmente a pergunta; \\
3 - altamente relevante - o enunciado responde a pergunta, tratando completamente de  suas nuances. \\
Em seguida, explique a razão para a escolha do escore. \\
Por favor, responda no formato JSON, contendo as chaves Razão e Score; \\
o valor de Razão deve ser a motivação para a escolha do score; \\
o valor de Score deve ser o valor do score atribuído. \\
\\
\textbf{[User]}\\
Pergunta: \{QUERY\}\\
Enunciado de jurisprudência: \{ENUNCIADO\}

\end{tcolorbox}
\caption{Prompt used to extract the relevance score.}
\label{fig:prompt_score_relevancia}
\end{figure}

\begin{figure}[!htb]
\begin{tcolorbox}[style_prompt]
\textbf{[System]}\\
Você é um especialista em sistemas de busca que usam o algoritmo BM25 e está trabalhando na indexação de uma base de dados de jurisprudência do Tribunal de Contas da União. Essa base está sofrendo com o problema de descasamento de vocabulário, ou seja, o usuário usa termos de pesquisa que não estão no enunciado da jurisprudência. Trata-se de um problema comum, pois o usuário não sabe como o enunciado está escrito.\\
\\
Para mitigar esse problema, você lerá um enunciado e escolherá as cinco palavras mais relevantes desse enunciado. Em seguida, escolherá dois ou três sinônimos para cada palavra.\\
\\
O formato de sua resposta deve ser:\\
- [Palavra 1]: [Sinônimos]\\
- [Palavra 2]: [Sinônimos]\\
- [Palavra 3]: [Sinônimos]\\
- [Palavra 4]: [Sinônimos]\\
- [Palavra 5]: [Sinônimos]\\
\\
Tudo o que você responder será indexado. Por isso, forneça apenas a lista das cinco palavras mais relevantes e seus sinônimos. Não inclua nenhuma instrução ou explicação sobre sua resposta.\\
\\
\textbf{[User]}\\
\{ENUNCIADO\}

\end{tcolorbox}
\caption{Prompt to extract synonyms for relevant words from the SUMMARY.}
\label{fig:prompt_synonym_expansion}
\end{figure}

\FloatBarrier

\section{Translated prompts into English}
\label{sec:prompts_en}

This appendix presents the translated version of the prompts shown in Appendix \ref{sec:prompts_pt}.

\begin{figure}[!htb]
\centering
\begin{tcolorbox}[style_prompt]
\textbf{[User]}

Create up to 5 short and direct questions that can be answered based on the following statement:\\

\{SUMMARY\}

\end{tcolorbox}
\caption{Prompt used to generate queries of groups G2 and G3 (translation of Figure \ref{fig:prompt_g2_g3}).}
\label{fig:prompt_g2_g3_en}
\end{figure}

\begin{figure}[!htb]
\begin{tcolorbox}[style_prompt]
\textbf{[System]}\\
You are an expert in the jurisprudence of the Federal Court of Accounts with the objective of evaluating whether a jurisprudential statement answers a question.\\
You must return a score value from 0 to 3, as follows:\\
0 - irrelevant - the statement does not answer the question; \\
1 - related - the statement is only on the topic of the question; \\
2 - relevant - the statement partially answers the question; \\
3 - highly relevant - the statement answers the question, fully addressing all its nuances. \\
Then, explain the reason for the score choice. \\
Please respond in JSON format, containing the keys Reason and Score; \\
the value of Reason should be the rationale for the choice of the score; \\
the value of Score should be the assigned score value. \\
\\
\textbf{[User]}\\
QUESTION: \{QUERY\}\\
Summary of the jurisprudence: \{SUMMARY\}

\end{tcolorbox}
\caption{Prompt used to extract the relevance score (translation of Figure \ref{fig:prompt_score_relevancia}).}
\label{fig:prompt_score_relevancia_en}
\end{figure}

\begin{figure}[!htb]
\begin{tcolorbox}[style_prompt]
\textbf{[System]}\\
You are an expert in search systems that use the BM25 algorithm and are working on indexing a database of jurisprudence from the Federal Court of Accounts. This database is facing the vocabulary mismatch problem, meaning the user uses search terms that are not in the statement of the jurisprudence. This is a common problem, as the user does not know how the statement is written.\\
\\
To mitigate this problem, you will read a statement and choose the five most relevant words from it. Then, you will select two or three synonyms for each word.\\
\\
The format of your response should be:\\
- [Word 1]: [Synonyms]\\
- [Word 2]: [Synonyms]\\
- [Word 3]: [Synonyms]\\
- [Word 4]: [Synonyms]\\
- [Word 5]: [Synonyms]\\
\\
Everything you respond with will be indexed. Therefore, provide only the list of the five most relevant words and their synonyms. Do not include any instructions or explanations about your answer.\\
\\
\textbf{[User]}\\
\{SUMMARY\}

\end{tcolorbox}
\caption{Prompt to extract synonyms for relevant words from the SUMMARY. (translation of Figure \ref{fig:prompt_synonym_expansion}).}
\label{fig:prompt_synonym_expansion_en}
\end{figure}

\FloatBarrier

\clearpage

\section{Queries}
\label{sec:anexo_queries}

\begin{longtable}{|c|l|}
    \caption{Query group G1.}
    \label{tab:G1}\\
    \hline
    \textbf{ID} & \textbf{Query} \\ \hline
    1  & técnica e preço \\
    2  & restos a pagar \\
    3  & aditivo a contrato \\
    4  & adesão a ata de registro de preços \\
    5  & sobrepreço e superfaturamento \\
    6  & restrição a competitividade \\
    7  & acréscimos e supressões \\
    8  & obras e serviços de engenharia \\
    9  & fiscalização de contratos \\
    10 & diárias e passagens \\
    11 & bens e serviços comuns \\
    12 & parcelas de maior relevância e valor significativo \\
    13 & despesa sem cobertura contratual \\
    14 & decreto-lei 4.657/1942 \\
    15 & contas e materialidade \\
    16 & inexequibilidade e comprovação \\
    17 & impedimento de licitar e contratar \\
    18 & aditivo e obra \\
    19 & fraude a licitação \\
    20 & auditoria interna \\
    21 & fracionamento de despesas \\
    22 & novo e improrrogável prazo \\
    23 & inidoneidade de licitante \\
    24 & licitações e contratos \\
    25 & exigência de atestado de capacidade \\
    26 & contrato com a administração pública \\
    27 & inexigibilidade e singularidade \\
    28 & mera participação e epp \\
    29 & decisão judicial \\
    30 & fiscal de contrato \\
    31 & medida cautelar \\
    32 & é possível a aplicação concomitante \\
    33 & reajuste de contrato \\
    34 & publicidade e concurso público \\
    35 & falecimento e multa \\
    36 & independência das instâncias \\
    37 & planejamento estratégico \\
    38 & sistema s e licitação \\
    39 & citação e validade \\
    40 & multa a particulares \\
    41 & licitação e preço de mercado \\
    42 & edital modificação \\
    43 & padronização marca \\
    44 & interesse recíproco \\
    45 & pesquisa de preços \\
    46 & planilha de custos e formação de preços \\
    47 & publicidade e propaganda \\
    48 & processo seletivo e sistema s \\
    49 & modalidade de licitação \\
    50 & antecipação de pagamento \\ \hline
\end{longtable}

\begin{longtable}{|c|p{0.95\linewidth}|}
    \caption{Query group G2.}
    \label{tab:G2}\\
    \hline
    \textbf{ID} & \textbf{Query} \\ \hline
    51 & concessão remunerada de uso de bens públicos modalidade \\ 
    52 & citação válida falecimento \\
    53 & garantia contratual patrimônio líquido mínimo \\
    54 & garantia de participação patrimônio líquido mínimo \\
    55 & despesas sem cobertura contratual multa \\
    56 & uso de áreas comerciais em aeroportos pregão \\
    57 & cessão das áreas comerciais de centrais públicas de abastecimento de gêneros alimentícios licitação \\
    58 & capacidade bens pertinentes e compatíveis com o objeto da licitação comprovação \\
    59 & extrapolação dos limites para alterações consensuais qualitativas de contratos de obras e serviços \\
    60 & responsabilidade do gestor sucessor omissão do antecessor \\
    61 & projeto de parceria público-privada modalidade \\
    62 & objeto do certame contrato social licitante \\
    63 & responsabilidade gestão dos recursos Fundo Municipal de Saúde \\
    64 & pesquisa de mercado orçamento fornecedores \\
    65 & pesquisa de preços obrigatória em licitações \\
    66 & vantajosidade adesão à ata \\
    67 & compensação entre acréscimos e supressões contratos administrativos \\
    68 & vedado restabelecimento item suprimido \\
    69 & marca aquisição cartuchos \\
    70 & SUS recursos União competência do TCU \\
    71 & preços referenciais sistemas oficiais estimativa de custos \\
    72 & ponderação técnica e preço \\
    73 & pontuação desarrazoada licitações técnica e preço \\
    74 & prorrogação de contrato administrativo término prazo de vigência \\
    75 & atestado de capacidade técnica serviços advocatícios conselho de fiscalização profissional \\
    76 & reposição de importâncias indevidamente percebidas boa-fé servidores erro escusável por órgão/entidade \\
    77 & aquisição direta risco a saúde \\
    78 & pregão divulgação dos preços estimados no edital \\
    79 & instrutor treinamento inexigibilidade de licitação \\
    80 & reajuste de preços e repactuação \\
    81 & recomposição reajuste \\
    82 & retenção de pagamentos perda da regularidade fiscal \\
    83 & irregulares despesas desnecessárias e anteriores à celebração do contrato \\
    84 & vínculo trabalhista ou societário pré-existente competitividade das licitantes \\
    85 & técnica e preço profissionais quadro permanente \\
    86 & erro material desclassificação da proposta \\
    87 & vedação à inclusão documento habilitação diligência \\
    88 & exigência comprovação de quantitativos mínimos características semelhantes \\
    89 & extensão sanção de impedimento de licitar e contratar \\
    90 & habilitação técnico-operacional obras e serviços de engenharia \\
    91 & controle adesão \\
    92 & alvará de funcionamento habilitação jurídica \\
    93 & acumulação de cargo secretário municipal professor \\
    94 & prescrição da pretensão punitiva julgamento das contas \\
    95 & Banco de Preços em Saúde referência medicamentos \\
    96 & alteração contratual serviços já previstos no edital \\
    97 & licitação princípio da legalidade estrita afastado \\
    98 & renovação de contrato serviços de natureza continuada pesquisa de preços \\
    99 & contratações de software cartas de exclusividade \\
    100 & assistência médica a servidores \\ \hline
\end{longtable}

\begin{longtable}{|c|p{0.95\linewidth}|}
    \caption{Query group G3.}
    \label{tab:G3}\\
    \hline
    \textbf{ID} & \textbf{Query} \\ \hline
    101 & Qual é a modalidade de licitação adequada para a concessão remunerada de uso de bens públicos? \\
    102 & A citação é considerada válida após o falecimento do responsável se a defesa já foi apresentada? \\
    103 & A prestação de garantia contratual é permitida juntamente com a exigência de patrimônio líquido mínimo? \\
    104 & A exigência de garantia de participação é permitida juntamente com a exigência de patrimônio líquido mínimo? \\
    105 & Quais são as consequências para os responsáveis que realizam despesas sem cobertura contratual? \\
    106 & O pregão é adequado para concessões de uso de áreas comerciais em aeroportos? \\
    107 & Quais normas devem ser observadas na cessão das áreas comerciais de centrais públicas de abastecimento de gêneros alimentícios? \\
    108 & É necessário exigir ao licitante comprovar que já forneceu bens pertinentes e compatíveis com o objeto da licitação? \\
    109 & É permitido a extrapolar os limites estabelecidos no art. 65, §§ 1º e 2º, da Lei 8.666/1993 para alterações consensuais qualitativas de contratos de obras e serviços? \\
    110 & Qual a responsabilidade do gestor sucessor quando o gestor antecessor omitiu o dever de prestar contas? \\
    111 & Qual é a modalidade de licitação utilizada para contratar serviços técnicos necessários à estruturação de projeto de parceria público-privada relativo a infraestrutura de rede de iluminação pública? \\
    112 & O que acontece se não houver compatibilidade entre o objeto do certame e as atividades previstas no contrato social da empresa licitante? \\
    113 & Quem é responsável pela gestão dos recursos do Fundo Municipal de Saúde? \\
    114 & É obrigatória pesquisa de mercado na elaboração do orçamento-base da licitação? \\
    115 & A pesquisa de preços correntes no mercado é obrigatória em licitações? \\
    116 & O que fazer para verificar a vantajosidade da adesão à ata? \\
    117 & É permitida a compensação entre acréscimos e supressões nos contratos administrativos? \\
    118 & Quais são as condições para que o restabelecimento total ou parcial de quantitativo de item anteriormente suprimido por aditivo contratual não configure compensação vedada pela jurisprudência do TCU? \\
    119 & A Administração pode indicar preferência por marcas em licitações para aquisição de cartuchos de tinta? \\
    120 & O TCU é responsável pela fiscalização das ações e serviços de saúde pagos com recursos repassados pela União no âmbito do Sistema Único de Saúde? \\
    121 & O que deve ser feito se não for possível obter preços referenciais nos sistemas oficiais para estimativa de custos em processos licitatórios? \\
    122 & É necessária ponderação entre a pontuação técnica e a de preço em uma licitação do tipo técnica e preço? \\
    123 & Como a pontuação desarrazoada pode limitar a competitividade nas licitações do tipo técnica e preço? \\
    124 & É possível prorrogar um contrato administrativo fora do término do prazo de vigência? \\
    125 & Qual é o problema na exigência de atestado de capacidade técnica para contratação de serviços advocatícios por conselho de fiscalização profissional? \\
    126 & É necessária reposição de importâncias indevidamente percebidas, de boa-fé, por servidores, em virtude de erro escusável de interpretação de lei por parte do órgão/entidade? \\
    127 & É possível aquisição direta quando a falta de produto ou serviço pode colocar em risco a saúde das pessoas? \\
    128 & É obrigatória a divulgação dos preços estimados no edital nos pregões para aquisição de medicamentos? \\
    129 & A contratação de professores, conferencistas ou instrutores para ministrar cursos de treinamento ou aperfeiçoamento de pessoal se enquadra na inexigibilidade de licitação? \\
    130 & Qual é a diferença entre reajuste de preços e repactuação? \\
    131 & É possível a aplicação da recomposição mesmo após a aplicação do reajuste previsto no contrato? \\
    132 & A retenção de pagamentos é permitida em caso de perda da regularidade fiscal? \\
    133 & Por que despesas desnecessárias e anteriores à celebração do contrato são consideradas irregulares? \\
    134 & Como o vínculo trabalhista ou societário pré-existente afeta a competitividade das licitantes no certame licitatório? \\
    135 & É permitido atribuir pontuação a uma empresa licitante nas licitações de técnica e preço com base na posse de determinados tipos de profissionais em seu quadro permanente? \\
    136 & Um erro material deve levar à desclassificação antecipada da proposta? \\
    137 & Há vedação à inclusão de novo documento destinado a atestar condição de habilitação preexistente em sede de diligência? \\
    138 & A exigência de comprovação de quantitativos mínimos em obras ou serviços com características semelhantes é legal? \\
    139 & Qual a extensão da sanção de impedimento de licitar e contratar? \\
    140 & Quais documentos são exigidos para habilitação técnico-operacional em certames de obras e serviços de engenharia? \\
    141 & Qual a finalidade do controle das autorizações de adesão? \\
    142 & Em que situações a apresentação do alvará de funcionamento é permitida na habilitação jurídica? \\
    143 & É permitida a acumulação do cargo de secretário municipal com o cargo de professor? \\
    144 & A prescrição da pretensão punitiva do TCU afeta o julgamento das contas? \\
    145 & O Banco de Preços em Saúde é uma referência de preços válida para a aquisição de medicamentos? \\
    146 & É possível a inclusão de serviços já previstos no edital em uma alteração contratual? \\
    147 & O princípio da legalidade estrita pode ser afastado em favor de outros princípios no procedimento licitatório? \\
    148 & Quais fontes devem ser prioritárias na pesquisa de preços para a renovação do contrato de serviços de natureza continuada? \\
    149 & Devem ser aceitas nas contratações de software as cartas de exclusividade emitidas pelos próprios fabricantes? \\
    150 & Qual é a regra geral para a contratação de entidade para prestação de serviços de assistência médica a servidores? \\
    \hline
\end{longtable}

\end{document}